\documentstyle[epsfig]{mn}

\title{A Simple Algorithm for Orbit Classification}

\author[E.E. Fulton and J.E. Barnes]{Eliza E. Fulton and
Joshua E. Barnes\thanks{E-mail: barnes@ifa.hawaii.edu}\\
        Institute for Astronomy, 2680 Woodlawn Dr., Honolulu, HI 96822, USA}

\begin{document}

\maketitle

\begin{abstract}
We describe a simple algorithm for classifying orbits into orbit
families.  This algorithm works by finding patterns in the sign
changes of the principal coordinates.  Orbits in the logarithmic
potential are studied as an application; we classify orbits into
boxlet families, and examine the influence of the core radius on the
set of stable orbit families.
\end{abstract}


\section{Introduction}

The dynamical structure of an elliptical galaxy may be represented by
a collection of orbits (e.g.~Schwarzschild 1979).  In this
representation the time-average of each orbit yields an invariant
density distribution, and a weighted sum of such distributions equals
the overall mass density of the galaxy.  Orbits are categorized into
different families depending on their shapes; certain orbit families
become stable or dominant in certain potentials.  The orbits available
in a given potential may be characterized in several interrelated
ways.  Classifying orbits into families says something about their
geometry and thus about the contribution they can make to the mass
distribution.  This paper describes a simple algorithm for classifying
orbits in elliptical potentials, and illustrates its application to
the well-studied logarithmic potential (see Miralda-Escud\'{e} \&
Schwarzschild 1989, hereafter MES89).

The orbital content of triaxial potentials has long been of interest
to dynamical astronomers.  Integrable potentials have fairly simple
structures, with one family of `box' orbits and three of `tubes'
(de~Zeeuw 1985).  In nonintegrable potentials the situation is much
more complex; chaotic orbits may appear (Goodman \& Schwarzschild
1981; Valluri \& Merritt 1998), and the monolithic tube and box
families are partly replaced by a plethora of minor orbit families
(Binney \& Spergel 1982; MES89; Lees \& Schwarzschild 1992).  This
rich structure is created by a pattern of resonances between
fundamental orbital frequences.  In $d$ dimensions there are $d$ such
frequencies $\omega_i$, and an orbit family corresponds to a resonance
condition of the form
\begin{equation}
    {\bf n} \cdot {\bf \bomega} = 0 \, ,
\end{equation}
where ${\bf n}$ is a vector of integers, and $|{\bf n}| > 0$; thus
each orbit family may be characterized by the corresponding vector
${\bf n}$.

Several different approaches to orbit classification exist.  Spectral
methods (Binney \& Spergel 1982, 1984; Papaphilippou \& Laskar 1996,
1998; Carpintero \& Aguilar 1998; Valluri \& Merritt 1998) classify
orbits by measuring their fundamental frequencies $\omega_i$.  These
methods are powerful and general but also somewhat complex to
implement.  Alternately, orbits may be classified by heuristic methods
based on known characteristics of specific orbit families.  For
example, tube orbits maintain a definite sense of circulation about a
principal axis (e.g.~Barnes 1992).  Likewise, the `boxlet' orbits
found in nonintegrable potentials may be classified by measuring
extrema along the principal coordinates (e.g.~Schwarzschild 1993).

We have developed a classification algorithm that captures the
geometry of both loop and boxlet orbits in a simple unified scheme.
This algorithm is easily implemented and generalized to find
higher-order resonances.  In its present form the algorithm is best
suited for two-dimensional potentials; while it provides a partial
characterization of orbits in three-dimensional potentials, it cannot
recognise the full range of orbit families present in such systems.
Nonetheless, the generality and simplicity of our method may make it
attractive when a full-blown spectral treatment is unnecessary or
prohibitively expensive.

In this paper we describe the algorithm and present some sample
results.  Section~2 develops the theoretical basis for the algorithm
and discuss some aspects of its implementation.  Section~3 illustrates
the use of this algorithm in exploring the orbits in the logarithmic
potential.  Conclusions and further applications of the algorithm are
discussed in Section 4.

\section{Algorithm}

The key idea of our algorithm is to determine an orbit's family from
the pattern of sign changes of the principal coordinates (Barnes 1998,
p.~352).  This approach can be placed on a firm foundation by
examining sign-change patterns for the closed, stable orbits which
parent an orbital family.  A given family actually contains a
one-parameter sequence of stable closed orbits, each with a different
energy.  In what follows we will use a single one of these closed
orbits as representative of an entire family.  This simplifies our
exposition but has no other consequences, since all the closed orbits
in a given family have the same topology and therefore the same
sign-change pattern.

\subsection{Closed orbits}

In two dimensions, the resonance condition (1) establishes a unique
relationship between the two fundamental frequencies of an orbit.  We
write (1) in the form
\begin{equation}
    \ell \omega_{\rm x} + m \omega_{\rm y} = 0 \, ,
\end{equation}
where $\ell$ and $m$ are integers, and $\omega_{\rm x}$ and
$\omega_{\rm y}$ are the orbital frequencies.  We assume with no loss
of generality (i) that the frequencies obey $0 < \omega_{\rm x} \le
\omega_{\rm y}$, (ii) that $m < 0 < \ell$, (iii) that $m$ and $\ell$
are relatively prime, and (iv) that the orbit has a period of $2 \pi$.
The resonance condition then implies that $\omega_{\rm x} = -m$ and
$\omega_{\rm y} = \ell$.

Since $\omega_{\rm x} / \omega_{\rm y}$ is a rational number, an orbit
with two frequencies $\omega_{\rm x}$ and $\omega_{\rm y}$ will close
on itself after a finite time.  Such a closed orbit, if stable, can
parent an orbital family.  Although the motion is generally not
harmonic, these closed orbits are topologically equivalent to
Lissajous figures, which are parametric curves generated by
\begin{equation}
    x(t) = \cos(\omega_{\rm x} t) \, , \qquad
    y(t) = \cos(\omega_{\rm y} t + \phi) \, .
\end{equation}
where $\phi \equiv k\pi / 2\omega_{\rm x}$ is a phase-shift.  For any
given $\omega_{\rm x}$ and $\omega_{\rm y}$ there are two values of
the parameter $k$ which yield possible parenting orbits: $k = 0$
generates the parents of boxlets, while $k = 1$ generates the parents
of `antiboxlets' (MES89).

Using the parametric equations (3), we can derive the pattern of sign
changes of $x(t)$ and $y(t)$ for any distinct choice for $(\ell,m)$
and $k$.  Since $\cos(\theta)$ has {\it two\/} sign changes for
$\theta \in [0,2\pi)$, this pattern is periodic with a period of
$\pi$, or half the orbital period.  We refer to the pattern generated
over the interval $t \in [0,\pi)$ as the {\it semi-pattern\/}; in each
semi-pattern $x(t)$ has $-m$ sign changes and $y(t)$ has $\ell$ sign
changes.  Before presenting a general treatment, we will discuss the
semi-patterns of some key orbits, organized by their $\ell + m$
values.  These examples also serve to introduce our notation for
sign-change patterns.

\subsubsection{The case $\ell + m = 0$}

If $\ell = -m$ then since $\ell$ and $m$ are relatively prime we have
$(\ell,m) = (1,-1)$, so $\omega_{\rm x} = \omega_{\rm y} = 1$.  In
this case, the curve defined by (3) with $k = 0$ is trivial: a
diagonal line passing through the origin.  On the other hand, the
curve produced with $k = 1$ is a circle, which is topologically
equivalent to a loop orbit.  For $t \in [0,\pi)$ the function $y(t)$
changes sign at $t = 0$, while $x(t)$ changes sign at $t = \pi/2$.
Abstracting away the actual values of $t$ but preserving the order of
events, this orbit's semi-pattern is represented by the string `YX',
where `Y' marks a sign-change of $y(t)$ and `X' a sign-change of
$x(t)$.  An orbit which produces a sign-change string of the form
`YX' repeated {\it ad infinitum\/} is a loop orbit.

\subsubsection{The case $\ell + m = 1$}

If $\ell = 1 - m$ and $k = 0$ then (3) yields `centrophobic' orbits
(MES89), which avoid the centre of the potential.  The semi-pattern
for such orbits is a string of the form `YX{\dots}Y', in which $-m$
pairs of the symbols `YX' are followed by a final `Y'.  For
example, the resonance $(\ell,m) = (2,-1)$ yields the semi-pattern
`YXY', and an orbit which produces a sign-change string consisting
of infinite repetitions of `YXY' is a centrophobic $(2,-1)$
resonance, also known as a `banana' orbit (Binney 1982).  Likewise,
the resonance $(3,-2)$ yields `YXYXY', which identifies a `fish'
orbit.  Unlike the loop orbit described above, all semi-patterns of
this kind must begin {\it and\/} end with the symbol `Y'.  The
sign-change string generated by an orbit of this kind will therefore
contain places where the `Y' appears twice with no intervening
`X'.  These doubled symbols have a simple physical interpretation:
the angular momentum of the orbit reverses between the first and
second `Y'.

In contrast, $k = 1$ yields `centrophilic' orbits, which pass exactly
through the centre of the potential.  The simplest such orbit is the
`antibanana', produced by $(\ell,m) = (2,-1)$.  For this orbit $y(t)$
has zeros at $t = 0$ and $\pi/2$, while $x(t)$ has a zero at $t =
\pi/2$.  To represent the simultaneous zeros of $x(t)$ and $y(t)$ at
$t = \pi/2$, we surround the symbols with square brackets; thus the
semi-pattern for this orbit is `Y[XY]'.  Next is the `antifish',
produced by $(\ell,m) = (3,-2)$; the semi-pattern for this orbit is
`YXY[XY]'.  In all cases with $\ell + m = 1$, the semi-patterns for
the centrophilic orbits resemble those of their centrophobic
counterparts, except that the final symbols are `[XY]', indicating
that the orbit passes through the origin.

\subsubsection{The case $\ell + m = 2$}

If $\ell = 2 - m$ then $k = 0$ yields centrophilic orbits which pass
through the origin at times just before $t = \pi/2$; thus the
semi-patterns for such orbits contain a central `[XY]' pair.  The
simplest orbit of this kind with $\omega_{\rm x}/\omega_{\rm y} \ge
1/2$ is $(\ell,m) = (5,-3)$, which produces the semi-pattern
`YXY[XY]YXY'.  The next resonance in this series, $(\ell,m) = (7,-5)$,
yields `YXYXY[XY]YXYXY'.  In general, each such pattern has three
parts: the first is an alternating string of $(\ell-1)/2$ copies of
`Y' and $(-m-1)/2$ copies of `X', the second is the central `[XY]'
pair, and the third is identical to the first.

For $k = 1$ we obtain centrophobic orbits.  Each semi-pattern contains
at total of $\ell$ copies of `Y' symbols and $\ell - 2 = -m$ copies
of `X'; thus within the semi-pattern two `Y' symbols must appear
next to each other.  The simplest example is $(\ell,m) = (5,-3)$,
which yields `YXYYXYXY'.  As already noted, `YY' pairs indicate a
reversal of the angular momentum; counting the two `Y' symbols which
begin and end the semi-period, we find that a centrophobic orbit with
$\ell + m = 2$ reverses its sense of angular momentum four times
before closing on itself.

\subsubsection{General treatment}

The coordinate functions $x(t)$ and $y(t)$ have zeros in the interval
$t \in [0,\pi)$ at times
\begin{equation}
\begin{array}{ll}
  \displaystyle
    \tau_{\rm x} = \frac{\pi}{-m} \left( i_{\rm x} + \frac{1}{2} \right) \, ,
    & \qquad i_{\rm x}  = 0, 1, \dots, -m-1 \, , \\
  \vspace{-0.2cm} \\
  \displaystyle
    \tau_{\rm y} = \frac{\pi}{\ell}
		     \left( i_{\rm y} + \frac{1}{2} + \frac{k}{2m} \right) \, ,
    & \qquad i_{\rm y}  = 0, 1, \dots, \ell-1 \, .
\end{array}
\end{equation}
Using these expressions, it's straightforward to generate the
semi-pattern for any given $(\ell,m)$ and $k$.  Table~1 gives results
for all resonances satisfying $\omega_{\rm x}/\omega_{\rm y} \ge 1/2$
which yield semi-patterns of $20$ symbols or less.  The resonances are
grouped by $\ell + m$ value; within each group, they are sorted in
order of increasing $\omega_{\rm x} / \omega_{\rm y}$; finally we list
the centrophobic version of each resonance before its centrophilic
counterpart.  The column labeled `Key' gives a one-character symbol
for each orbit family, which we will use in presenting results of
orbit classification in Section~3.

\begin{table}
\caption{Semi-patterns of $20$ symbols or less with $\omega_{\rm
x}/\omega_{\rm y} \ge 1/2$.}
\begin{tabular}{rrccl}
$\ell$ &  $m$ & $k$ & Key & Pattern \\
\vspace{-0.2cm} \\
   $1$ & $-1$ & $1$ &   L & YX \\
   $1$ & $-1$ & $0$ &   l & [XY] \\
\vspace{-0.2cm} \\
   $2$ & $-1$ & $0$ &   b & YXY \\
   $2$ & $-1$ & $1$ &   B & Y[XY] \\
   $3$ & $-2$ & $0$ &   f & YXYXY \\
   $3$ & $-2$ & $1$ &   F & YXY[XY] \\
   $4$ & $-3$ & $0$ &   p & YXYXYXY \\
   $4$ & $-3$ & $1$ &   P & YXYXY[XY] \\
   $5$ & $-4$ & $0$ &   q & YXYXYXYXY \\
   $5$ & $-4$ & $1$ &   Q & YXYXYXY[XY] \\
   $6$ & $-5$ & $0$ &   r & YXYXYXYXYXY \\
   $6$ & $-5$ & $1$ &   R & YXYXYXYXY[XY] \\
   $7$ & $-6$ & $0$ &   s & YXYXYXYXYXYXY \\
   $7$ & $-6$ & $1$ &   S & YXYXYXYXYXY[XY] \\
   $8$ & $-7$ & $0$ &   t & YXYXYXYXYXYXYXY \\
   $8$ & $-7$ & $1$ &   T & YXYXYXYXYXYXY[XY] \\
   $9$ & $-8$ & $0$ &   u & YXYXYXYXYXYXYXYXY \\
   $9$ & $-8$ & $1$ &   U & YXYXYXYXYXYXYXY[XY] \\
  $10$ & $-9$ & $0$ &   v & YXYXYXYXYXYXYXYXYXY \\
  $10$ & $-9$ & $1$ &   V & YXYXYXYXYXYXYXYXY[XY] \\
\vspace{-0.2cm} \\
   $5$ & $-3$ & $1$ &   @ & YXYYXYXY \\
   $5$ & $-3$ & $0$ &   2 & YXY[XY]YXY \\
   $7$ & $-5$ & $1$ &  \# & YXYXYYXYXYXY \\
   $7$ & $-5$ & $0$ &   3 & YXYXY[XY]YXYXY \\
   $9$ & $-7$ & $1$ &  \$ & YXYXYXYYXYXYXYXY \\
   $9$ & $-7$ & $0$ &   4 & YXYXYXY[XY]YXYXYXY \\
  $11$ & $-9$ & $1$ &  \% & YXYXYXYXYYXYXYXYXYXY \\
  $11$ & $-9$ & $0$ &   5 & YXYXYXYXY[XY]YXYXYXYXY \\
\vspace{-0.2cm} \\
   $7$ & $-4$ & $0$ &   w & YXYYXYXYYXY \\
   $7$ & $-4$ & $1$ &   W & YXYYXY[XY]YXY \\
   $8$ & $-5$ & $0$ &   x & YXYXYYXYYXYXY \\
   $8$ & $-5$ & $1$ &   X & YXY[XY]YXYYXYXY \\
  $10$ & $-7$ & $0$ &   y & YXYXYYXYXYXYYXYXY \\
  $10$ & $-7$ & $1$ &   Y & YXYXYYXYXY[XY]YXYXY \\
  $11$ & $-8$ & $0$ &   z & YXYXYXYYXYXYYXYXYXY \\
  $11$ & $-8$ & $1$ &   Z & YXYXY[XY]YXYXYYXYXYXY \\
\vspace{-0.2cm} \\
   $9$ & $-5$ & $1$ &  \& & YXYYXYYXYXYYXY \\
   $9$ & $-5$ & $0$ &   7 & YXYYXY[XY]YXYYXY \\
  $11$ & $-7$ & $1$ &   * & YXYXYYXYYXYXYYXYXY \\
  $11$ & $-7$ & $0$ &   8 & YXYXYYXY[XY]YXYYXYXY \\
\vspace{-0.2cm} \\
  $11$ & $-6$ & $0$ &   g & YXYYXYYXYXYYXYYXY \\
  $11$ & $-6$ & $1$ &   G & YXYYXYYXY[XY]YXYYXY \\
  $12$ & $-7$ & $0$ &   h & YXYYXYXYYXYYXYXYYXY \\
  $12$ & $-7$ & $1$ &   H & YXYYXYXYYXYYXY[XY]YXY \\
\vspace{-0.2cm} \\
  $13$ & $-7$ & $1$ &   ? & YXYYXYYXYYXYXYYXYYXY \\
  $13$ & $-7$ & $0$ &   0 & YXYYXYYXY[XY]YXYYXYYXY \\
\end{tabular}
\end{table}

In passing, we note a crucial fact: no semi-pattern consists of
multiple copies of any other semi-pattern.  This follows because
$\ell$ and $m$ are relatively prime.

The motion described by (3) is {\it reversible\/}; that is, the
transformation $t \to -t$ leaves trajectories unchanged.  Thus it may
seem puzzling that some semi-patterns are not palindromes.  The
resolution of this puzzle lies in the fact that the transformation $t
\to t + \Delta t$, where $\Delta t$ is a constant, also leaves
trajectories unchanged.  Consequently, all cyclic permutations of a
given semi-pattern should be identified with each other since they all
represent the same trajectory.  The reverse of a given semi-pattern
can always be cyclicly permuted to obtain the original semi-pattern.
We conclude that the asymmetric appearance of some semi-patterns is an
artifact of our arbitrary choice for the origin of the time-line.

\subsection{Family resemblances}

The closed orbits in elliptical potentials are distorted versions of
Lissajous figures, and each such figure yields a unique semi-pattern.
We now argue that all the orbits in an orbital family yield similar
sign-change patterns.  Recall that in two dimensions, each orbit
family is parented by a {\it stable\/} closed orbit.  Stability is the
key which enables an orbit to be a parent; informally, it implies that
other orbits which start near the parenting orbit remain near it at
later times.  An orbital family is thus the set of all orbits which
remain near a stable parenting orbit.

For a centrophobic family the resemblance is simple; all the orbits in
a given family generate the {\it same\/} semi-pattern as does their
parenting orbit.  This seems intuitively clear for orbits which wander
only a small distance from the path of their parent, but is it true
for even the most far-flung family members?  Yes, because no member of
a centrophobic family passes exactly through the centre.  Consider a
continuous ensemble of family members which -- up until now -- have
all generated the same sequence of $x$ and $y$ sign-changes, and are
therefore at present all within the same $(x,y)$ quadrant of the
system.  If some of these orbits next have sign-changes in $x$ while
others have sign-changes in $y$ then the swath defined by their orbits
must pass across the origin.  But continuity then implies the ensemble
includes orbits which pass exactly through the centre, which is
impossible since the family is centrophobic.

On the other hand, a centrophilic family includes at least one member
-- the parenting orbit -- which passes exactly through the centre
every time.  Other family members come arbitrarily close to the
centre, but generally pass to one side or the other; the swath defined
by an ensemble of such orbits has finite width as it crosses the
origin.  Thus some of members of the ensemble have sign-changes in the
order `XY' and some in the order `YX', while only a infinitesimal
subset pass exactly through the centre and produce `[XY]'.  The
sign-change sequence generated by a typical centrophilic family member
is an infinite repetition of its parenting orbit's semi-pattern with
one embellishment -- where the parent's semi-pattern contains `[XY]'
the member's realization may contain either `XY' or `YX' (but not `XX'
or `YY').  Thus, each `[XY]' indicates a place where the order of the
symbols is interpreted as {\it ambiguous\/}.

This interpretation complicates matters, since the semi-pattern of a
centrophobic resonance will match the semi-pattern of its centrophilic
counterpart.  Nonetheless, an orbit can be assigned to its appropriate
family with little ambiguity -- given a sufficiently long sequence of
sign-change data generated by following the orbit's trajectory.  Match
the orbital data against each semi-pattern in turn, taking care to
test centrophobic resonances before testing centrophilic ones; the
first successful match gives the orbit's family.  With a finite
sequence of sign-change data, it's possible that a centrophilic
resonance will be miss-classified as centrophobic, but the probability
of such an error is small if enough trajectory data is available.

\subsection{Implementation}

A fairly straightforward computer program serves to implement the
algorithm outlined above.  This program computes the trajectory
$(x(t),y(t))$ of an orbit, recording the sign-changes of $x$ and $y$;
once a total of $N_{\rm change}$ sign-changes have been recorded,
these are matched against a finite set of semi-patterns.  A successful
match identifies the orbit's family.  If none of the semi-patterns
match the orbit's sign-change data then the latter are searched for
{\it any\/} strictly periodic pattern of length $\le L_{\rm max}$;
this catches all centrophobic families with $\ell-m \le L_{\rm max}$.
If no such pattern is found then the orbit may be a higher-order
resonance, a true (regular) box with incommensurate $\omega_{\rm x}$
and $\omega_{\rm y}$, or a stochastic orbit; our algorithm can't
distinguish between these possibilities.

The trajectory-following section of the program is built around a
$4^{\rm th}$-order RK integrator with a time-step always proportional
to the local dynamical time.  While a higher-order integrator and
local error control would improve the accuracy of the computed
trajectories, convergence tests indicate that most orbits can be
robustly classified even when integrated with energy errors as large
as $0.1$ per cent; to be on the safe side, we limit the peak-to-peak
energy variation to $0.02$ per cent.  After each time-step, the new
values of $x$ and $y$ are compared with their previous values; if one
has changed sign then this event is recorded, while if both have
changed sign then both events are recorded, with linear interpolation
used to determine which occurred first.  The algorithm does not try to
detect passages exactly through the origin; such passages are
vanishingly rare.

The semi-patterns used to try and match the sign-change data are
generated from the $(\ell,m)$ and $k$ values stored in a table like
Table~1.  When matching the orbital data to a semi-pattern we demand
symbol-by-symbol equality throughout unless the semi-pattern contains
`[XY]'; at that point the orbital data may contain either `XY' or
`YX'.  One final subtlety about the matching routine is worth
mentioning: since the phase of the orbital data is indeterminate, it
is matched against all cyclic permutations of each semi-pattern.  Even
with this added complication, the amount of computer time spent
matching orbital data to patterns is insignificant; only integer
arithmetic is required, and most possible matches are quickly
rejected.

Finally, the routine which checks the orbital data for arbitrary
periodic patterns is quite simple.  To test for a semi-pattern of
length $L$ we take the first $L$ symbols as the model and match them
against the rest of the data.  In this matching process it's not
necessary to consider cyclic permutations of the semi-pattern since it
automatically has the same phase as the orbit.  If the match succeeds,
the $\ell$ and $m$ values of the parenting resonance are determined by
counting the number of `Y' and `X' symbols, respectively, in the
semi-pattern.  If the match fails then the routine next checks for
semi-patterns of length $L + 1, \dots, L_{\max}$.

The availability of this final `safety-net' for arbitrary periodic
patterns poses an interesting question: do orbit integrations ever
yield periodic sign-change sequences which can't be generated by the
scheme in Section~2.1?  No evidence for such orbits has been found.
For example, Table~1 contains all semi-patterns of length $L \le 20$
which can be generated by the scheme.  Among the many thousands of 2-D
orbits studied in the next section we have found {\it none\/} with
periodic semi-patterns of length $L \le 20$ which are not in Table~1.
This success illustrates the power of MES89's {\it ansatz\/} linking
periodic orbits to Lissajous figures.

\section{Application}

To demonstrate the use of our algorithm we made a survey of orbits in
the logarithmic potential,
\begin{equation}
    \Phi = \frac{1}{2} \ln \left( R_{\rm c}^2 + x^2 +
				    \frac{y^2}{b^2} \right) \, ,
\end{equation}
with ellipticity $b \le 1$ and core radius $R_{\rm c}$ (Richstone
1980; Binney \& Tremaine 1987; MES89).  The goal of this survey was to
show how the core radius influences the set of orbit families.  We
focused on orbits of total energy
\begin{equation}
    E \equiv \Phi({\bf r}) + \frac{1}{2} |\dot{\bf r}|^2 = 0 \, .
\end{equation}
This restriction to $E = 0$ is largely a matter of convenience.  In a
singular logarithmic potential ($R_{\rm c} = 0$) an orbit with $E \ne
0$ can always be rescaled to one with $E = 0$.  If $R_{\rm c} \ne 0$
then rescaling is still possible if the core radius changes along with
the energy so that $E\,-\,\ln R_{\rm c}$ stays constant.  Thus fixing
$E$ and changing $R_{\rm c}$, as we do in Section~3.2, is equivalent
to fixing $R_{\rm c}$ and changing $E$, and no new orbital families
would come to light if we were to vary {\it both\/} parameters instead
of just one.

In designing an orbit survey, perhaps the most critical choice is the
`start space' which samples the manifold of possible orbits.  Below
we examine several possible choices.  First, we present results for a
two-dimensional start space which is guaranteed to include all orbits
but awkward to use when surveying a large number of potentials.
Second, we present results for several one-dimensional start spaces
which between them allow access to all known orbits.

\subsection{2-D start space}

The phase space for a two-dimensional potential has four dimensions:
two positions, $(x, y)$, and two velocities, $(\dot{x}, \dot{y})$.
Since all orbits eventually cross the $y$ axis, we stipulate that the
initial $x$ coordinate is $x_0 = 0$.  The equation for $E$ above then
fixes any one of the remaining coordinates in terms of the other two.
We constructed a grid in $y_0$ and $\dot{y}_0$, and found $\dot{x}_0$
using (6).  The resulting two-dimensional start space is parameterized
by the same coordinates often used when constructing a `surface of
section' (H\'{e}non \& Heiles 1964; Contopoulos 1983; Binney \&
Tremaine 1987, p.~117).

\begin{figure*}
\begin{center}
\epsfig{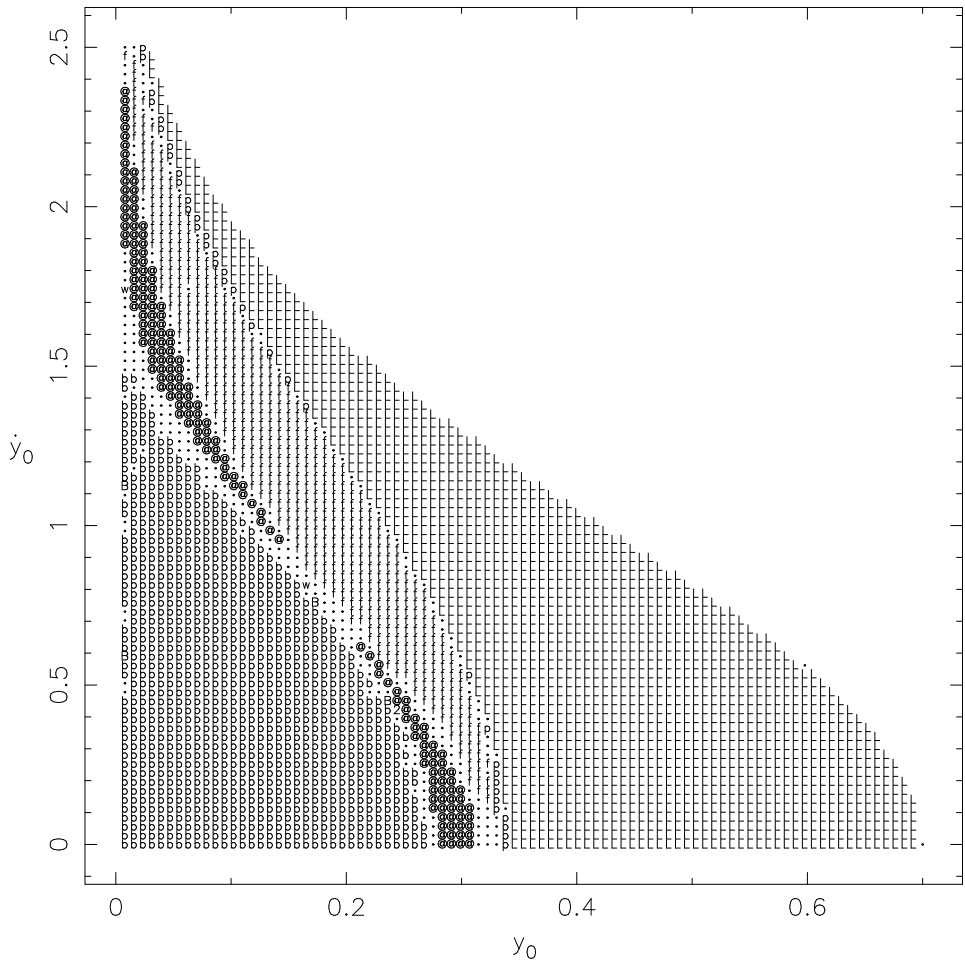}
\end{center}
\caption{Orbit families accessible from the 2-D start space.  The
potential has core radius $R_{\rm c} = 0$ and ellipticity $b = 0.7$.
The symbol plotted at each grid point indicates the classification
according to the `Key' in Table~1.}
\end{figure*}

\begin{figure*}
\begin{center}
\epsfig{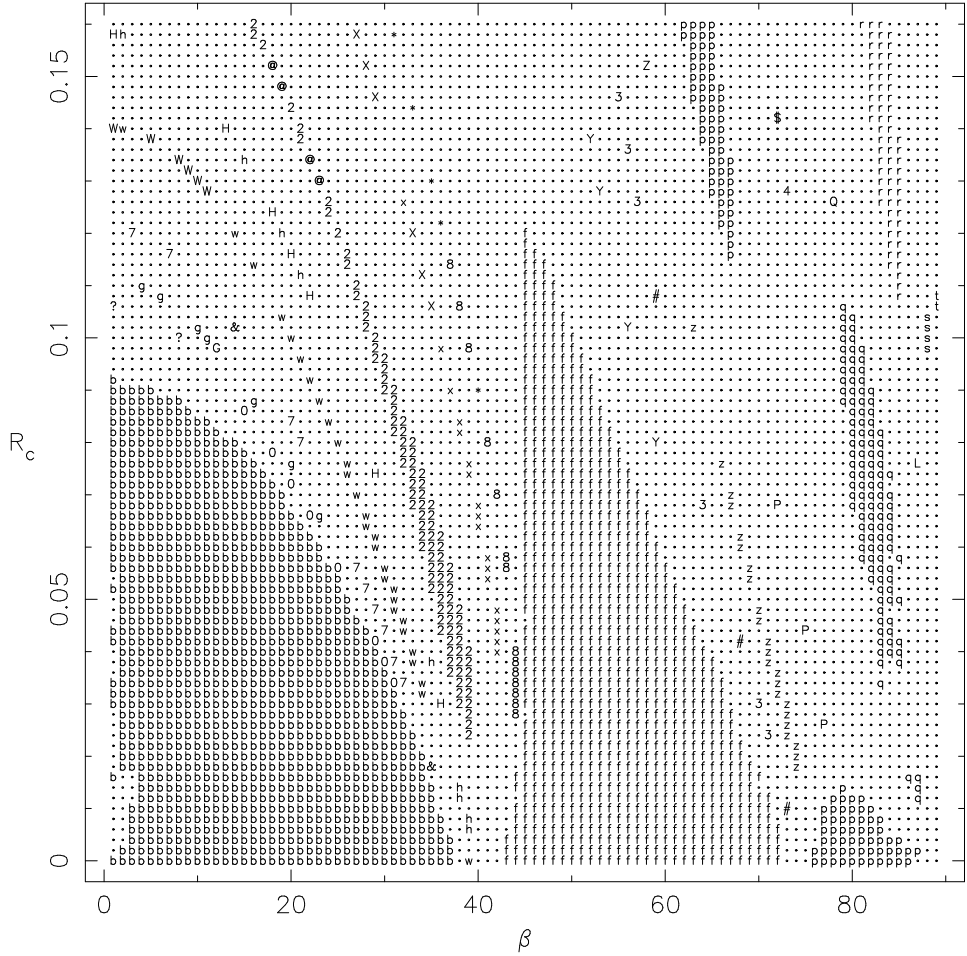}
\end{center}
\caption{Orbit families accessible from the stationary start space.
The potential has ellipticity $b = 0.8$.  The vertical axis shows
$R_{\rm c}$, the core radius, while the horizontal axis shows $\beta$,
the angular coordinate of the initial position.}
\end{figure*}

\begin{figure*}
\begin{center}
\epsfig{figure=fig03_80.ps,width=5.0in}
\end{center}
\caption{Orbit families accessible from the central start space.
The potential has ellipticity $b = 0.8$.  The vertical axis shows
$R_{\rm c}$, the core radius, while the horizontal axis shows
$\beta'$, the angular coordinate of the initial velocity.}
\end{figure*}

\begin{figure*}
\begin{center}
\epsfig{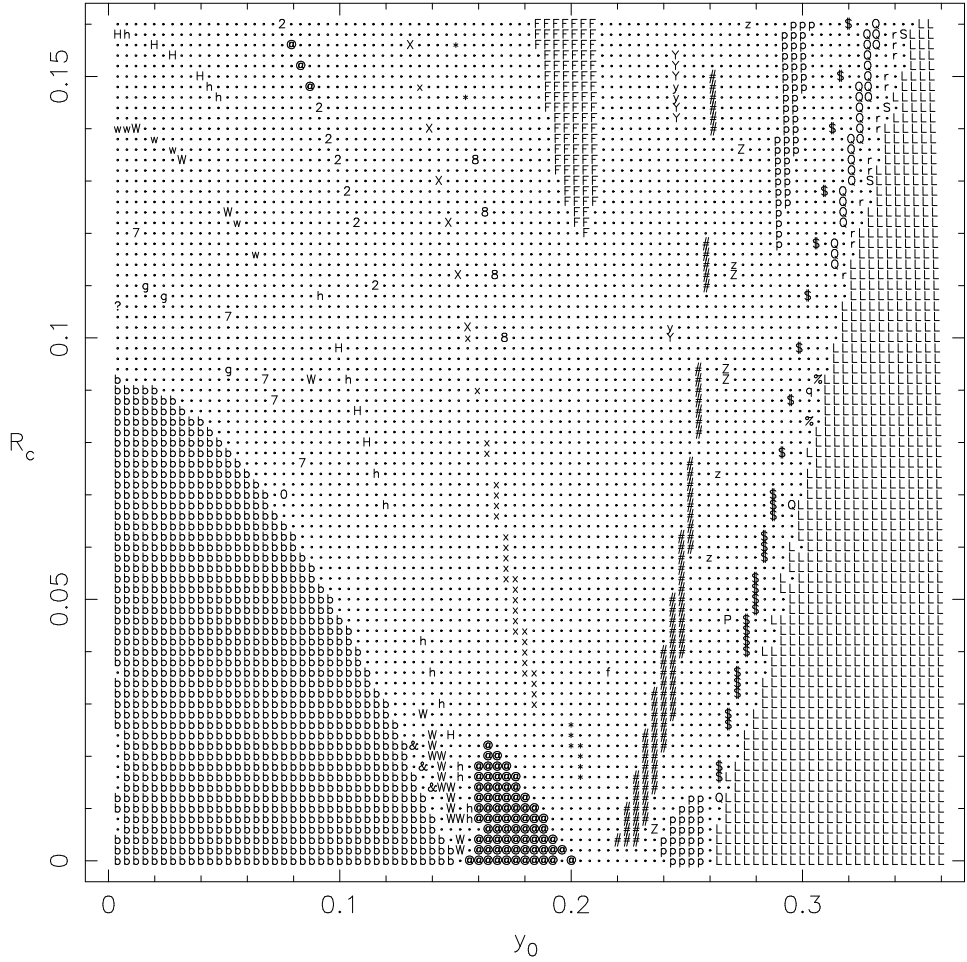}
\end{center}
\caption{Orbit families accessible from the perpendicular start space.
The potential has ellipticity $b = 0.8$.  The vertical axis shows
$R_{\rm c}$, the core radius, while the horizontal axis shows $y_0$,
the initial position along the minor axis.}
\end{figure*}

Fig.~1 shows the result of classifying orbits in the logarithmic
potential with core radius $R_{\rm c} = 0$ and ellipticity $b = 0.7$
using this two-dimensional start space.  We emphasise that this
diagram is {\it not\/} a surface of section; rather than plotting an
orbit's position on the $(y, \dot{y})$ plane each time it intersects
$x = 0$, we plotted the results of classifying separate orbits
launched from each point on the grid.  We use the symbols listed in
the `Key' column of Table~1 to indicate the classifications, with a
small dot marking orbits which don't belong to any family with $\ell -
m \le 20$.  Our result may be directly compared to those of other
orbit-classification schemes (Carpintero \& Aguilar 1998, Fig.~16) and
to the traditional surface of section for this potential (MES89,
Fig.~1); the agreement is very good.  Basically, the phase space for
this potential is almost entirely occupied by a few centrophobic orbit
families.  Starting at the lower left we find a large number of
$(\ell,m) = (2,-1)$ banana orbits.  Moving diagonally across the plot,
we next encounter a strip, broken in the middle, of $(5,-3)$
resonances.  Next comes a region of $(3,-2)$ fish orbits, followed by
a narrow strip of $(4,-3)$ pretzel orbits.  Finally, the right-hand
side of the plot is entirely occupied by $(1,-1)$ loop orbits.

The classifications shown in Fig.~1 were made using $N_{\rm change} =
1000$ sign-changes per orbit.  This gives a high level of confidence;
for example, every fish orbit in this diagram has executed $200$
semi-patterns without a single sign-change out of place.  Since the
orbit families in this particular potential all have fairly short
semi-patterns, we could have obtained results almost as good with less
than $100$ sign-changes per orbit.  A potential which supports a wider
range of resonances provides a more stringent test of the classifier.
We therefore chose a potential with $R_{\rm c} = 0.02$ and $b = 0.8$;
compared to the potential used for Fig.~1, the finite core radius
helps stabilize centrophilic orbits, while the more modest flattening
favors high-order resonances.  The same orbits were independently
classified for $N_{\rm change} = 125$, $250$, $500$, and $1000$, and
the results compared to determine which orbit classifications are
sensitive to this parameter.  Even with $N_{\rm change} = 125$, the
$(\ell,m) = (2,-1)$, $(5,-3)$, $(3,-2)$, and $(4,-3)$ resonances were
accurately classified, but incorrect classifications sometimes
resulted for higher resonances such as $(9,-5)$, $(7,-4)$, $(8,-5)$,
and $(7,-5)$.  We observed two kinds of classification errors: false
positives from non-resonant orbits which happen to lie {\it near\/}
stable resonances, and mistaken identities due to confusion between
centrophobic and centrophilic versions of the same resonance.  The
output of the classifier converges nicely as $N_{\rm change}$ is
increased; taking the results for $N_{\rm change} = 1000$ as fiducial,
we find total error rates of $4.3$, $1.9$, and $0.8$ per cent, for
$N_{\rm change} = 125$, $250$, and $500$, respectively.

\subsection{1-D start spaces}

In most previous studies of boxlets the initial conditions were
generated by choosing initial positions ${\bf r}_0$ on the
equipotential surface $\Phi({\bf r}_0) = E$ and setting the initial
velocities $\dot{\bf r}_0$ to zero.  This `stationary' start space
yields a wide variety of boxlets, but it excludes orbits like the
antifish which don't have a stationary point -- that is, a point at
which $|\dot{\bf r}| = 0$.  Alternatively, if the potential is finite
at the origin one may consider setting the initial positions ${\bf
r}_0$ to zero and choosing initial velocities satisfying $|\dot{\bf
r}_0| = (2E - 2\Phi(0))^{1/2}$ (Papaphilippou \& Laskar 1996).  This
`central' start space includes the antifish, but it excludes
centrophobic orbits.  When using the stationary start space we
parameterize the initial position in terms of the angle $\beta$
between ${\bf r}_0$ and the $x$ axis; likewise when using the central
start space we parameterize the initial velocity in terms of the angle
$\beta'$ between $\dot{\bf r}_0$ and the $x$ axis.


Fig.~2 presents a survey of the stationary start space for logarithmic
potentials with ellipticity $b = 0.8$ and core radii $R_{\rm c}$ in
the range $0.16$ to $0.0$.  As in Fig.~1, the result of each orbit
classification is shown using the symbols listed in Table~1.  For the
larger values of $R_{\rm c}$ boxlets are few and far between, with
only the $(\ell,m) = (4,-3)$ and $(6,-5)$ resonances occupying more
than a handful of orbits.  As $R_{\rm c}$ decreases, low order
resonances such as $(2,-1)$ and $(3,-2)$ appear and grow in
importance, eventually dominating the start space as $R_{\rm c} \to
0$.  The general trend with core radius seen here is entirely
consistent with the results presented by MES89.

Although the overall outline of Fig.~2 is unsurprising, some of the
higher-order resonances present in this diagram may be less familiar.
For example, there is a narrow band of centro{\it philic\/} $(\ell,m)
= (5,-3)$ orbits extending from the top of the diagram and ending near
$R_{\rm c} \simeq 0.02$ and $\beta \simeq 39^\circ$.  This family is
distinct from the centrophobic $(5,-3)$ resonance seen in Fig.~1 and
in previous studies (MES89; Lees \& Schwarzschild 1992).  In addition,
this centrophilic family is flanked on either side by two centrophobic
higher-order resonances: the $(7,-4)$ and $(8,-5)$ families.  Only the
former has been noted in the literature (Lees \& Schwarzschild 1992),
but the overall pattern seen here might have been anticipated since
rational numbers, each corresponding to a possible resonance, are
dense among the reals.  The pattern of resonant orbits in Fig.~2 may
be summarized as follows: at some starting angle $\beta$ between any
two resonances, $(\ell_1,m_1)$ and $(\ell_2,m_2)$, one may look for a
resonance of the form $(\ell_1 + \ell_2, m_1 + m_2)$.  Of course, not
all resonances are stable, and unstable resonances don't parent orbit
families.  Nonetheless, we expect that much of the space in Fig.~2
occupied by unclassified orbits (dots) is actually threaded by a fine
array of still higher-order families.


But Fig.~2 can't capture all the stable resonances; it only shows
orbits which have a stationary point.  In Fig.~3, constructed using a
central start space, we partly remedy this deficiency; this plot
includes {\it all\/} centrophilic orbits, both those that have
stationary points and those that do not.  We find a number of new
orbit families.  For larger values of $R_{\rm c}$ there is a region of
$(\ell,m) = (3,-2)$ orbits (antifish) centred on $\beta' \simeq
45^\circ$, and a narrow strip of $(5,-4)$ orbits near $\beta' \simeq
78^\circ$.  At intermediate values of $R_{\rm c}$ there is a fairly
wide region of $(4,-3)$ resonances (antipretzels) near $\beta' \simeq
75^\circ$.  Finally, the bottom part of the plot contains regions of
$(7,-4)$ resonances near $\beta' \simeq 18^\circ$ and $(8,-5)$
resonances near $\beta' \simeq 47^\circ$.  These families, and several
others, are not represented in Fig.~2.

Together, Figs.~2 and~3 exhibit an interesting regularity; as $R_{\rm
c}$ changes, a given boxlet vanishes from one plot at almost the same
point where the corresponding antiboxlet appears in the other, and
{\it vice versa\/}.  For example, the centrophilic $(\ell,m) = (3,-2)$
family is confined to $R_{\rm c} \ga 0.12$, while the centrophobic
version of the same family is confined to $R_{\rm c} \la 0.12$.  A
similar statement applies to the pair of $(4,-3)$ resonances and to
other pairs of resonances as well.  This behavior can be understood
from MES89's discussion of bifurcations in the logarithmic potential;
they remark that as a function of $R_{\rm c}$, `a boxlet and its
antiboxlet change stability in such a manner that generally one is
stable and the other is unstable'.


Does this rule extend to all resonances?  The centrophilic $(\ell,m) =
(5,-3)$ family is seen in both Figs.~2 and~3 for $R_{\rm c} \ga 0.02$,
but neither plot shows the complementary family.  This is easily
explained; the missing family is centrophobic and has no stationary
point, so it's automatically excluded from both plots.  We therefore
constructed a `perpendicular' start space by setting $x_0 =
\dot{y}_0 = 0$ and using (6) to fix $\dot{x}_0$ in terms of $y_0$.
This start space allows access to centrophobic orbits which don't have
a stationary point -- provided that each orbit at some point crosses
the $y$ axis at an angle of $90^\circ$.  Fig.~4 shows the results of
classifying orbits from this start space for $y_0$ in the range
$0.004$ to $0.36$.  (We excluded the axial orbit obtained with $y_0 =
0$, and trimmed off the loop orbits which dominate this start space
from $y_0 \simeq 0.3$ out to the maximum value, $y_{\rm max} = b (1 -
R_{\rm c}^2)^{1/2} \simeq 0.8$.)  This new plot shows the missing
centrophobic $(5,-3)$ family appearing as expected for $R_{\rm c} \la
0.02$; it's centred on $y_0 \simeq 0.18$.  Several other families in
Fig.~4 also reinforce this pattern; for example, the small clutch of
centrophobic $(11,-7)$ orbits centred on $y_0 \simeq 0.2$, $R_{\rm c}
\simeq 0.02$ neatly complements the narrow bands of centrophilic
$(11,-7)$ orbits which extend down to $R_{\rm c} \simeq 0.03$ in
Figs.~2 and~3.

Fig.~4 also shows several families previously undetected.  The
centrophobic $(\ell,m) = (7,-5)$ resonance occupies a strip extending
from $y_0 \simeq 0.22$ at the bottom of the plot to $y_0 \simeq 0.26$
at the top.  This strip has appreciable width for smaller core radii
but narrows to less than the space between adjacent grid points in
$y_0$ for $R_{\rm c} \ga 0.08$, thereby creating apparent gaps where
the strip misses the grid.  A similar effect is seen in the
centrophobic $(9,-7)$ family.  Both of these families seem stable for
almost all -- if not quite all -- values of $R_{\rm c}$ tested, so
it's probably not worth looking for their centrophilic counterparts.

\section{Discussion}

Our algorithm works by reducing orbital trajectories to sequences of
symbols, and matching these sequences against a set of known patterns.
The idea of describing trajectories symbolically is not new; it has
been applied to geodesics on negatively-curved surfaces, to forced van
der Pol oscillators, and to the restricted three-body problem (Moser
1973 and references therein).  Symbolic representation of orbits may
have other applications in galactic dynamics.  Below we briefly
discuss a few related issues and some possible future developments.

\subsection{Start spaces}

The reader may feel somewhat reassured to learn that the three 1-D
start spaces considered above are probably both necessary and {\it
sufficient\/} to find all boxlet families.  We can show this by using
the properties of the parametric curves (3).  First, the orbits which
have stationary points are those for which $k = 0$.  Second,
centrophilic orbits are those for which the quantity $\ell + m + k$ is
an even number.  Third, orbits which cross the $y$ axis at an angle of
$90^\circ$ are those for which $\ell + 2m + k$ is even.  Now if an
orbit has no stationary point and is centrophobic then it has $k = 1$
and $\ell + m + k$ is odd, and because $\ell$ and $m$ are relatively
prime it follows that $m$ is odd and $\ell + 2m + k$ is even -- so the
orbit crosses the $y$ axis at $90^\circ$.  Thus {\it any\/} orbit must
fall into at least one of these three categories.  Orbits with $k = 0$
can be accessed from the stationary start space, orbits with even
$\ell + m + k$ are reachable using the central start space, and orbits
with even $\ell + 2m + k$ result from the perpendicular start space.
With a few exceptions due to misclassification of high-order
resonances, these rules accurately predict which orbit families appear
in each of Figs.~2, 3, and~4.

The stationary and perpendicular start spaces suffice for the singular
logarithmic potential ($R_{\rm c} = 0$) since all centrophilic orbits
are unstable in this potential (e.g.~Kuijken 1993).

These results have some bearing on orbit surveys in three-dimensional
potentials.  At a given energy, the manifold of possible orbits can be
covered by a 4-D start space constructed by analogy with the 2-D start
space described in Section~3.1 (Levison \& Richstone 1987).  Since
this start space is awkward to sample and to display, many workers
favor 2-D start spaces analogous to the stationary and perpendicular
start spaces described above (Schwarzschild 1993).  But these 2-D
start spaces don't allow access to centrophilic orbits which lack a
stationary point, and these orbits may be stable in nonsingular
potentials.  Thus a full survey of orbits in a three-dimensional
nonsingular potential requires {\it at least\/} three different 2-D
start spaces: the stationary and perpendicular start spaces must be
supplemented by a central start space (Merritt 1999).  It still
remains to be shown that these 2-D spaces really are sufficient to
access all orbits.

\subsection{Noninteger $k$?}

We've assumed, following MES89, that the {\it only\/} allowed values
of the phase parameter are $k = 0$ and $k = 1$.  This assumption is
critically important for the discussion in the previous subsection.
However, D.~Merritt (private communication) has pointed out that $k$
may have values between $0$ and $1$; for example, in a 2-D harmonic
oscillator the $x$ and $y$ motions are independent and neutrally
stable orbits with {\it any\/} real value of $k$ exist.  For
nonintegrable potentials the situation is unclear; $k$ is probably
limited to a discrete set of values, but it's possible that values
between $0$ and $1$ are permitted.  Our algorithm can't distinguish a
resonant orbits with noninteger $k$ from a centrophobic orbit with the
same $(\ell,m)$, since both yield the same sign-change pattern.  Such
orbits would be excluded from the 1-D start spaces used in
Section~3.2, though in principle they are accessible from a 2-D start
space.  Visual inspection may be the easiest way to discover orbits
with noninteger $k$; such orbits, while following the same sign-change
patterns as `normal' centrophobes, would have different symmetry
properties.  No such orbits have been seen in a cursory inspection of
candidates, but more work is needed.

\subsection{Orbit families}

The results presented in Section~3 emphasize the remarkable variety
and structure of orbit families in the logarithmic potential.  Of the
$46$ families listed in Table~1, some $29$ can be found in Figs.~2, 3,
and~4.  Almost half of these families are centrophilic.  While
centrophilic orbits have been described before (MES89; Lees \&
Schwarzschild 1992; Barnes 1998, Fig.~42), most earlier studies
treated these families as isolated curiosities.  MES89's bifurcation
diagrams showed that boxlets and antiboxlets exchange parental roles
as $R_{\rm c}$ changes; we have made the consequences of this
role-reversal visible by routinely finding the centrophilic
counterparts of unstable centrophobic orbits.  Centrophilic orbits may
be destabilized by central mass concentrations in real galaxies, but
their recognition still helps in understanding the mechanisms which
determine the orbital content of a potential.

We note that some families did not appear in any of our classification
diagrams.  Most of these missing families are fairly high-order
resonances which may be difficult to find because they occupy very
small regions of the start spaces.  Two low-order centrophilic
families, however, are conspicuously absent.  One is the orbit
generated by the $(\ell,m) = (1,-1)$ resonance with $k = 0$, and the
other is the antibanana, generated by the $(2,-1)$ resonance with $k =
1$.  The former resonance is stable in some anharmonic potentials (de
Zeeuw \& Merritt 1983), while the latter may occur in highly flattened
logarithmic potentials (Papaphilippou \& Laskar 1996, Fig.~10).  We
would not expect these centrophilic orbits to appear in our plots
since their centrophobic counterparts, the loop and banana orbits, do
appear and are stable for a wide range of parameters.

\subsection{3-D orbits}

It's somewhat disappointing to admit that the technique described here
doesn't trivially generalize to three-dimensional potentials.  The
reason for this failure is fairly straightforward.  In two dimensions
the resonance condition (1) fixes a unique relation between the
fundamental frequencies, but in three dimensions it does not; as a
result, the parenting orbits of many three-dimensional orbit families
are thin membranes rather than closed curves (Merritt \& Valluri
1999).  Such parenting orbits don't generate unique periodic
sign-change sequences, so our present algorithm can't recognise them.
In contrast, spectral methods work quite well in three dimensions
(Carpintero \& Aguilar 1998; Papaphilippou \& Laskar 1998; Valluri \&
Merritt 1998).

Nonetheless, we can partly classify orbits in three-dimensional
potentials by projecting their trajectories onto the principal planes
and using our two-dimensional algorithm to analyse the resulting
sign-change sequences.  This approach recognises resonances in which
one component of the integer vector ${\bf n}$ in (1) is zero; such
resonances, though not the rule, are fairly common in many potentials
(Papaphilippou \& Laskar 1998; Merritt \& Valluri 1999).  Some orbits
obey two independent resonance conditions; such `doubly-degenerate'
orbits are parented by closed three-dimensional curves.  But even
doubly-degenerate orbits don't always yield strictly periodic
sign-change sequences.  For example, the centrophobic
`banana-pretzel' orbit, combining a $(2,-1)$ resonance in the
$x$--$z$ plane with a $(4,-3)$ resonance in the $x$--$y$ plane, yields
a `ZYXZ[YZ]X[ZY]ZXYZ' semi-pattern containing not one but {\it
two\/} ambiguously-ordered pairs of symbols.  This example suggests
that more general pattern-matching methods may be able classify the
crossing-sequences generated by arbitrary three-dimensional
resonances.

\subsection{Future developments}

There are several ways in which this algorithm could be improved or
extended.  As just mentioned, one line of attack is to generalize the
pattern-scanning process.  It seems fairly easy to extend the routine
which searches for arbitrary periodic patterns (Section~3.2) to
recognise centrophilic as well as centrophobic semi-patterns; this
would provide a better safety-net for high-order resonances.  It's
also straightforward, in principle, to couple the trajectory-following
and pattern-matching routines together so that the latter can call the
former as needed for more data to resolve ambiguous cases.  This would
be more efficient and reliable than first integrating the orbit for a
fixed number of sign-changes and then analysing the results.

Another line of attack is to adapt the algorithm to analyse N-body
simulations.  A simplified version has already been used to classify
orbits in merger remnants (Barnes 1998; Bendo \& Barnes 2000);
however, this version just followed individual trajectories in a
frozen approximation of the N-body potential.  A more interesting
project would be to classify orbits directly from the computed
trajectories in a simulation.  This could be accomplished by using an
N-body code to generate separate sign-changes sequences for all $N$
bodies; these sequences could be quickly scanned by pattern-matching
algorithms like the one described here.  Such techniques could attach
a concise historical record to each body in an N-body simulation,
nicely complementing the instantaneous snapshots of positions and
velocities which currently form the basis for most N-body analysis.

\vspace{0.9cm} \noindent

We thank the referee, Dr.~P.~Palmer, for an open and constructive
report, and L.~Aguilar, T.~de Zeeuw, D.~Merritt, and E.~Roberts for
valuable and encouraging comments on a preliminary draft of this
paper.  This work was supported in part by NASA grant NAG 5-8393 and
STScI grant GO-06430.03-95A.

\end{document}